\documentstyle[12pt]{article}

\date{}

\begin{document}

\title{{\LARGE\sf Non-Mean-Field Behavior of Realistic Spin Glasses}}
\author{
{\bf C. M. Newman}\thanks{Partially supported by the
National Science Foundation under grants DMS-92-09053 and 95-00868.}\\
{\small \tt newman\,@\,cims.nyu.edu}\\
{\small \sl Courant Institute of Mathematical Sciences}\\
{\small \sl New York University}\\
{\small \sl New York, NY 10012, USA}
\and
{\bf D. L. Stein}\thanks{Partially supported by the
U.S.~Department of Energy under grant DE-FG03-93ER25155.}\\
{\small \tt dls\,@\,physics.arizona.edu}\\
{\small \sl Dept.\ of Physics}\\
{\small \sl University of Arizona}\\
{\small \sl Tucson, AZ 85721, USA}
}
\maketitle

{\bf Abstract:}  We provide rigorous proofs which show that the main
features of the Parisi solution of the Sherrington-Kirkpatrick spin glass
are not valid for more realistic spin glass models in any dimension and
at any temperature.

\bigskip
\bigskip

The theoretical perspective provided by the Parisi solution
\cite{Parisi1} of the
infinite-ranged Sherrington-Kirkpatrick (SK) model \cite{SK} has dominated
the spin glass literature over the past decade and a half.  This is partly
because it represents the only example of a reasonably complete thermodynamic
solution to an interesting and nontrivial spin glass model, and partly because
of the novel, and in some respects, spectacular, nature of the symmetry
breaking displayed in the low-temperature phase.  Its main qualitative
features --- the presence of (countably) many pure states,
the non-self-averaging of their
overlap distribution function, and the ultrametric organization of their
overlaps, among others --- have greatly influenced thinking about disordered
and complex systems in general \cite{BY,Parisi3}.  A common working hypothesis
is that the Parisi solution provides a theory of general
spin glass models \cite{BY,MPV,Parisi3}.  In particular,
many authors have directly applied its
features to the study of both short-ranged
models and laboratory spin glasses
\cite{Bouchaud92,Bouchaud94,Franz,Lederman}.  Support for this
``SK picture'' --- that the main qualitative features of
Parisi's solution survive in non-infinite-ranged
models --- comes from both analytical \cite {Georges} and
numerical \cite{Caracciolo,Reger} work.

In this Letter, however, we prove that
this approach is fundamentally flawed.
That is, short-ranged models such as the standard
nearest-neighbor Edwards-Anderson (EA) model \cite{EA}
cannot have
at {\it any\/} temperature all the basic features of the Parisi
solution.  Furthermore, most of our arguments
rely on
little more than the homogeneity properties of the disorder,
and thus are applicable to more realistic spin glass models
such as models with long-ranged couplings or diluted RKKY interactions
\cite{Note1}.

We do not attempt to resolve in this paper the closely related issue of
whether short-ranged spin glass
models have many pure thermodynamic states at sufficiently
high dimension and low
temperature, or only a single pair.  The latter conjecture arises from
a droplet model \cite{FH1}
based on a scaling {\it ansatz\/} \cite{Mac,BM,FH1}.
Rather, we assert that {\it if\/}
there are many pure states, their structure and
that of their overlaps cannot be that of the SK picture.

\medskip

{\it The SK picture.\/} --- The Parisi solution, when applied to the EA model
at fixed temperature $T$,
suggests that there exist two related quantities which are
non-self-averaging (i.e., depending on the realization
${\cal J}$ of the couplings):
(i) a state $\rho_{\cal J}(\sigma)$,
which is a Gibbs probability measure (at temperature $T$) on the microscopic
spin configurations $\sigma$ on all of $Z^d$, and (ii) a Parisi order parameter
distribution $P_{\cal J}(q)$, which is a probability measure on the interval
$[-1,1]$ of possible overlap values.  These two are related
as follows:  if one chooses $\sigma$ and $\sigma'$ from the
product distribution
$\rho_{\cal J}(\sigma)\rho_{\cal J}(\sigma')$,
then the overlap
\begin{equation}
\label{eq:overlap}
Q=\lim_{L\to\infty}|\Lambda_L|^{-1}\sum_{x\in\Lambda_L}\sigma_x\sigma'_{x}
\end{equation}
has $P_{\cal J}$ as its probability distribution.  Here $|\Lambda_L|$ is the
volume of a cube $\Lambda_L$ of side length $L$ centered at
the origin in $d$ dimensions.

A crucial component
of the SK picture is that the decomposition of
$\rho_{\cal J}$ into pure states is {\it countable\/}
(i.e., a sum rather than an integral):
\begin{equation}
\label{eq:sum}
\rho_{\cal J}(\sigma)=\sum_\alpha W_{\cal J}^\alpha\rho_{\cal J}^\alpha
(\sigma)\ .
\end{equation}
If $\sigma$ is drawn from $\rho_{\cal J}^\alpha$ and $\sigma'$
from $\rho_{\cal J}^\beta$, then the expression in Eq.~(\ref{eq:overlap})
equals
its thermal mean,
\begin{equation}
\label{eq:qab}
q_{\cal J}^{\alpha\beta}=\lim_{L\to\infty}|\Lambda_L|^{-1}
\sum_{x\in\Lambda_L} \langle\sigma_x\rangle_\alpha
\langle\sigma_x\rangle_\beta \quad .
\end{equation}
Thus $P_{\cal J}$ is given by
\begin{equation}
\label{eq:PJ(q)}
P_{\cal J}(q)=\sum_{\alpha,\beta}W_{\cal J}^\alpha W_{\cal J}^\beta
\delta(q-q_{\cal J}^{\alpha\beta})\quad .
\end{equation}

In the SK picture, the $W_{\cal J}^\alpha$'s and $q_{\cal J}^{\alpha\beta}$'s
are non-self-averaging quantities,
except for $\alpha=\beta$ or its global flip (where
$q_{\cal J}^{\alpha\beta}=\pm q_{EA}$).
The average $P(q)$ of
$P_{\cal J}(q)$ over the disorder
distribution $\nu$ of the couplings is a mixture of two delta-function
components at $\pm q_{EA}$ and a continuous part between them.

The countability of the decomposition of Eq.~(\ref{eq:sum})
is also employed to obtain
the often-used result (see, for example, Refs.~\cite{BY,Mezard84,Bouchaud92})
that the free energies of the lowest-lying states are independent random
variables with an exponential distribution.

Both $\rho_{\cal J}$ and $P_{\cal J}$ are infinite-volume quantities and
so must be obtained by some kind of thermodynamic limit.  Naively, one
might simply fix ${\cal J}$ and attempt to take a sequence of increasing
volumes with, say,
periodic boundary conditions.  However, we argued in a previous
paper \cite{NS92} that the existence of multiple pure states is
inconsistent with the existence of such a limit for {\it fixed\/}
${\cal J}$.
Instead, there would be chaotic size dependence,
so that infinite-volume limits can be achieved only through
coupling-{\it dependent\/} boundary conditions.  We will see
below that, nonetheless, $P_{\cal J}$ (and $\rho_{\cal J}$)
can be obtained by natural limit procedures
which are coupling-independent and which imply translation
invariance for $P_{\cal J}$ (and translation covariance for $\rho_{\cal J}$).
We ask whether this is consistent with the SK picture, which requires the
following properties of $P_{\cal J}$ and its average $P$:

\medskip

\noindent 1) $P_{\cal J}(q)$ is non-self-averaging.

\noindent 2) $P_{\cal J}(q)$ is a sum of (infinitely many) delta-functions.

\noindent 3) $P(q)$ has a continuous component (for all $q$ between
the delta-functions at $\pm q_{EA}$).

\medskip

The answer is no;  we will see that {\it
translation invariance rules out non-self-averaging\/}.
This in turn makes the absence of a continuous component in $P_{\cal J}$
inconsistent with its presence in $P$.
We conclude that {\it property 1) is absent,
and at most one of the remaining two properties
of the SK picture can be valid for realistic spin glass models\/}.  We
will consider below the implications of this result for other important
features of the SK picture, such as ultrametricity.

\medskip

{\it Construction of $\rho_{\cal J}$ and $P_{\cal J}$.\/} --- We first
describe a limit procedure to obtain $P_{\cal J}$ which does not
involve the prior construction of $\rho_{\cal J}$.  Begin with the finite
volume Gibbs distribution $\rho_{{\cal J}^{(L)}}^{(L)}$ on the spin
configuration $\sigma^{(L)}$ in the cube, $\Lambda_L$,
with periodic boundary conditions.  Here ${\cal J}^{(L)}$
denotes the couplings restricted to $\Lambda_L$.  Let $Q^{(L)}$ denote
the overlap of $\sigma^{(L)}$ and a duplicate $\sigma'^{(L)}$:
\begin{equation}
\label{eq:QL}
Q^{(L)}=|\Lambda_L|^{-1}\sum_{x\in\Lambda^{(L)}}\sigma_x^{(L)}
{\sigma'}_x^{(L)}\quad .
\end{equation}

The distribution $P_{{\cal J}^{(L)}}^{(L)}$ for $Q^{(L)}$ is the finite volume
Parisi overlap distribution function, whose average was studied numerically in
Refs.~\cite{Caracciolo,Reger}.
It was proved in Ref.~\cite{NS92} that in the SK model,
non-self-averaging requires $P^{(L)}_{{\cal J}^{(L)}}$ to have
chaotic $L$-dependence as $L\to\infty$ for fixed ${\cal J}$; a similar result
was suggested, though not proved, for short-ranged spin glasses with many
pure states.   Because of this, we do not take a limit of $P^{(L)}_{{\cal
J}^{(L)}}$
directly but rather of the {\it joint distribution\/} $\tilde\mu_L$ of
${\cal J}^{(L)}$ and $Q^{(L)}$.  That is, by a compactness argument (which
may require the use of a subsequence of $L$'s) one has a limiting
$\tilde\mu$, which is a probability measure on joint configurations
$({\cal J},q)$ ($q$ being a realization of $Q$) such that for
any (nice) function $f$ of {\it finitely\/} many couplings
and of $q$, the average $\langle f\rangle$ for $\tilde\mu$ is the limit
of the averages for $\tilde\mu_{L}$.

This gives us existence of a $\tilde\mu$, which is a joint distribution
on the infinite-volume realizations of ${\cal J}$ and $q$.  Its
marginal distribution for ${\cal J}$ is the original disorder distribution
$\nu$, while its conditional distribution for $q$ given ${\cal J}$ is
what we denote $P_{\cal J}$.  Because
of the periodic boundary conditions, the marginal distribution (under $\tilde
\mu_L$) of $J_1,\ldots,J_m,q$ is the same (for large $L$) as of $J_1^a,
\ldots,J_m^a,q$ (where $a$ is any lattice translation and ${\cal J}^a$ is the
translated ${\cal J}$) and thus one has
translation invariance of the limit measure $\tilde\mu$.
Translation invariance here means that for any $a$,
the shifted variables ${\cal J}^a$
together with $Q$ have the same joint distribution as do the
original ${\cal J}$ together with $Q$; because $\nu$
is in any case translation-invariant,
this implies that $P_{\cal J}=P_{{\cal J}^a}$.  In other words,
the overlaps don't care about the choice of origin.

Our second procedure for obtaining  $P_{\cal J}$ is first to construct
$\rho_{\cal J}$ and then obtain $P_{\cal J}$ as the distribution of the $Q$
given
by Eq.~(\ref{eq:overlap}).
The construction of
$\rho_{\cal J}$ is as follows.  Let $\mu_L$ be the joint
distribution for ${\cal J}^{(L)}$ and $\sigma^{(L)}$ on the periodic cube
$\Lambda_L$.  Then by compactness arguments, some
subsequence $\mu_{L}$ converges to a limiting joint distribution
$\mu({\cal J},\sigma)$.  The resulting conditional distribution of
$\sigma$ given ${\cal J}$ is what we denote $\rho_{\cal J}(\sigma)$.
$\mu$ will be translation-invariant (and $\rho_{\cal J}$ will be
translation-covariant) because of the translation
invariance (on the torus) of $\mu_L$.  Translation invariance
means that the distribution $\mu$ for
$({\cal J}, \sigma)$ is the same as for $({\cal J}^a,\sigma^a)$ for any
lattice vector $a$.  In terms of $\rho_{\cal J}$, this means that
$\rho_{{\cal J}^a}(\sigma)=\rho_{\cal J}(\sigma^{-a})$, so that, e.g.,
$\langle\sigma_x\rangle_{{\cal J}^a}=\langle\sigma_{x-a}\rangle_{\cal J}$;
thus we say that $\rho_{\cal J}$ is translation-covariant rather than
invariant.

Before pursuing the implications of translation invariance, we raise several
questions related to our constructions.  Could different subsequences
of cubes yield different limits?  We believe the answer is no,
although we have no complete proof, because our procedure of
considering {\it joint\/} distributions (for ${\cal J}$ and $q$ or
for ${\cal J}$ and $\sigma$) should eliminate the
kind of chaotic volume dependence
discussed in Ref.~\cite{NS92}.  Could different deterministic boundary
conditions yield different limits?  Boundary conditions related to each other
by partial or complete spin flips (e.g., periodic and antiperiodic) must have
the same limiting joint distributions (by arguments similar to those used
in Ref.~\cite{NS92}), but, in principle, unrelated boundary
conditions such as periodic and free could have different limits.  In practice,
however, we expect that
different sequences of deterministic boundary conditions would yield the same
(translation-invariant) limit.  Could the $P_{\cal J}$'s arising from our
two constructions (one using $\rho_{\cal J}$ and one not) be different?  We
shall take as a working hypothesis that the two are the same, but see no
compelling reason why that should be the case.  Either way,
since both $P_{\cal J}$'s are translation-invariant, neither one can be
non-self-averaging, as we now explain.

\medskip

{\it Self-averaging of $P_{\cal J}(q)$.\/} --- To justify our claim that
translation invariance of $P_{\cal J}(q)$ implies that it is
self-averaging, take a
(nice) function $f(q)$ (like $q^k$) and consider the function of ${\cal J}$,
$\hat f({\cal J})\equiv\int f(q)P_{\cal J}(q)\ dq$.  By translation
invariance, $\hat f({\cal J})=\hat f({\cal J}^a)$, but by the
{\it translation-ergodicity\/} \cite{Note2} of $\nu$,
any translation-invariant (measurable) function
$\hat f({\cal J})$ is equal to its ${\cal J}$-average, $\int\hat f({\cal J})
\nu({\cal J})\ d{\cal J}$.  Since this is true for all $f$'s, it
follows that $P_{\cal J}$ itself equals its ${\cal J}$-average.

We remark that the above discussion makes it clear that our claim is valid
for any model involving disorder whose underlying distribution is (like $\nu$)
translation-invariant and translation-ergodic.  For example,
any analogue of the Parisi
order parameter distribution for spin glass models with site-diluted RKKY
interactions will also be self-averaging (if it is translation-invariant).

Because $P_{\cal J}$ is self-averaging, we are forced to the dichotomy
that, for any temperature in any dimension, either $P(=P_{\cal J})$ is a sum
of one or more $\delta$-functions
or else $P$ has a continuous component.  When there is a unique
infinite-volume Gibbs state (e.g., in the paramagnetic phase)
then of course $\rho_{\cal J}$ is that state and $P$
is a single $\delta$-function at $q=0$.  If there were only two pure states
(related by a global flip) \cite{HF}, then $P$ would simply be a sum
of two $\delta$-functions at $\pm q_{EA}$.  But what
if infinitely many pure states $\rho_{\cal J}^\alpha$ coexist in
$\rho_{\cal J}$, with infinitely many overlap values $q_{\cal
J}^{\alpha\beta}$?
If the set of overlap values were {\it countably\/} infinite, then
$P_{\cal J}$ would necessarily be a sum of $\delta$-functions, but {\it the
infinitely many locations (as well as the weights) would not depend on
${\cal J}$\/}; we regard this possibility as implausible.

Thus we suggest that the most likely scenario for many coexisting pure states
and overlaps is one where the countable decomposition Eq.~(\ref{eq:sum}) is
replaced by an integral and where $P$ has no $\delta$-function
components.

\medskip

{\it Ultrametricity.\/} --- We briefly turn to the question of whether
ultrametricity of pure state overlaps \cite{Note3} can survive in
short-ranged spin
glasses, given that
$P_{\cal J}$ is self-averaged. Clearly, this type of nontrivial
ultrametricity requires the existence of multiple pure states.
As discussed above, we consider
the case where $P(q)$ is continuous.
We now demonstrate that such
an overlap distribution cannot have an ultrametric structure, in the Parisi
sense.

Let $\alpha,\beta,\gamma_1,\gamma_2\ldots$ denote pure states randomly selected
from the continuum of such states, and let their overlaps as
usual be denoted $q^{\alpha\beta}$, etc.  In the Parisi solution,
these overlaps are such that, for any $k$, the two
smallest of $q^{\alpha\beta},q^{\alpha\gamma_k}$, and $q^{\beta\gamma_k}$
are equal.  For nontrivial ultrametricity such as occurs in the
Parisi solution,
there would be positive probability that for some $i$ and $j$ the following
two strict inequalities occur simultaneously:
\begin{equation}
\label{eq:ultra}
q^{\alpha\gamma_i}<q^{\beta\gamma_i}\quad {\rm and}\quad
q^{\alpha\gamma_j}>q^{\beta\gamma_j}\quad .
\end{equation}
If ultrametricity holds, then the first inequality requires that
$q^{\alpha\gamma_i}=q^{\alpha\beta}$, while the second inequality
requires that $q^{\beta\gamma_j}=q^{\alpha\beta}$. Thus
$q^{\alpha\gamma_i}=q^{\beta\gamma_j}$.  But because $\alpha$, $\beta$,
$\gamma_i$, and $\gamma_j$ are chosen randomly and independently,
the two variables $q^{\alpha\gamma_i}=q^{\beta\gamma_j}$ are also
independent.  Because each of these is chosen from a {\it continuous\/}
distribution $P$, the probability that
the two overlap values can be identical is zero,
and we arrive at a contradiction.

The only way to avoid the contradiction is if the two strict inequalities
in Eq.~(\ref{eq:ultra}) {\it cannot\/} occur simultaneously.  This means that
either $q^{\alpha\gamma_k}\le q^{\beta\gamma_k}$ for {\it every\/} $k$ or
vice-versa, which implies that the pure states can be ordered into a
one-dimensional continuum, and the ultrametric structure resembles a comb
rather than a more usual tree, such as appears in the SK picture.

As discussed in the previous section (see also the next section),
self-averaging makes it implausible that the
set of overlaps is countable.  A countable set of overlaps
would invalidate the above argument and possibly rescue ultrametricity, but
at the cost of destroying anything resembling the Parisi solution.

\medskip

{\it Decomposition into pure states.\/} --- What is the nature of the
decomposition of $\rho_{\cal J}$ into pure states?  The SK picture
prediction
of a countably infinite sum as in Eq.~(\ref{eq:sum}) (with infinitely
many $q_{\cal J}^{\alpha\beta}$'s) has been
largely ruled out since the set of $q_{\cal J}^{\alpha\beta}$'s would be
self-averaging, which seems unreasonable.  Even if one were unwilling
to rule out countability on that ground, there are other arguments, not
presented here, which make that possibility even more unlikely.  These
arguments suggest that all $\rho_{\cal J}^{\alpha}$'s appearing in a countable
decomposition would have the same even spin correlations.  This certainly
seems inconsistent with the expected presence of domain walls between
pure states unrelated by a global spin flip.
We conclude that in any reasonable scenario for $\rho_{\cal J}$,
there should be at most one pair of pure states (related by a
global spin flip) with {\it strictly\/} positive weight.

In other words, either (a) $\rho_{\cal J}$ is pure, or
(b) it is a sum of two pure states related by a global flip, or
(c) it is an integral over pure states with none
having strictly positive weight, or (d) it has one
``special'' pair of pure states with strictly positive weight and all
the rest with zero weight.  Case (a) occurs
if the system is in a paramagnetic phase,
or any other in which the EA order parameter is zero.  Case (b)
would occur according to the Fisher-Huse droplet picture \cite{FH1}, but
could also occur if there existed multiple pure states
not appearing in $\rho_{\cal J}$ (``weak Fisher-Huse'') \cite{Note4}.
Case (c) occurs if there are
{\it uncountably\/} many pure states in the decomposition of $\rho_{\cal J}$,
all with zero weight (``democratic multiplicity'').  Case (d) (which
we regard as unlikely) occurs when one pair of pure states partially
dominates all others, but accounts for only part of the total weight
(``dictatorial multiplicity'').

What is the relation between the nature of $P(=P_{\cal J})$ and
the three (nontrivial) cases (b) -- (d) discussed above?
Clearly case (b) implies that $P$ is a sum of two $\delta$-functions at
$\pm q_{EA}$, and no continuous part.  If we assume in cases (c) and (d)
that varying $\alpha$ and $\beta$ through the continuous portion of
the pure states yields a continuously varying $q_{\alpha\beta}$
(but see the next paragraph for a case where this assumption
is violated), then it
follows that case (c) corresponds to a $P$ with {\it no\/} $\delta$-functions
while case (d) corresponds to a $P$ with $\delta$-functions at $\pm q_{EA}$
{\it and\/} a continuous part.  This latter case is the $P$ predicted by
the Parisi solution, but note two crucial distinctions between case (d) and
the SK picture:  (i) there is self-averaging, so one already obtains the
continuous part of $P$ from a single realization ${\cal J}$, and (ii) the
$\delta$-functions at $\pm q_{EA}$ come from a {\it single\/} special pair
of pure states --- not from countably many $q^{\alpha\alpha}$'s.
{\it Thus any numerical evidence in favor of such a distribution for\/}
$\left[P_{\cal J}\right]_{\rm av}$ (as in
Refs.~\cite{Caracciolo,Reger}) {\it is not
evidence in favor of the SK picture\/} but rather, at most,
supports dictatorial multiplicity.

We remark that a case of democratic multiplicity occurs in a
solution for the ground state structure in a short-ranged, highly
disordered spin glass model \cite{NS94}.
We argued there that below eight dimensions, there exists
a single pair of ground states (case (b) above), while above eight, there are
uncountably many.  It is not hard to see that the $\rho_{\cal J}$ for
$d>8$ corresponds to case (c)
above -- the states are chosen by the flips of fair coins for all the trees
in the invasion forest, so all have equal (zero) weight.  It appears that
here $P(q)$ is a $\delta$-function at zero!  So even for short-ranged
spin glasses with $T>0$, we cannot rule out the possibility of
democratic multiplicity with such a $P$.

\medskip

{\it Discussion and Conclusions.} ---  We have shown that in realistic
spin glass models, and probably all non-infinite-ranged spin glasses,
a natural construction leads to a Parisi overlap
distribution $P_{\cal J}$ which is translation-invariant and hence
self-averaging, unlike the Parisi solution of the SK model.  Although
our construction uses periodic boundary conditions, we believe that, in
short-ranged (and probably all non-infinite-ranged) models, any
choice of {\it coupling-independent\/} boundary conditions (e.g., free)
should yield the same translation-invariant $P_{\cal J}$.  Any
restoration of the SK picture would require a non-translation-invariant
$P_{\cal J}$ and we see no natural mechanism for obtaining one.  Even
were such a mechanism available, we note that the construction of our
self-averaged $P_{\cal J}$ shows that the SK picture can be, at best,
incomplete.  Any theory of the thermodynamics of realistic spin glasses
will likely differ considerably from the SK picture.

If many phases do exist in some dimension and below some temperature, we
believe that the most reasonable possibility is (c) above, i.e.,
democratic multiplicity.  If numerical experiments find an order
parameter distribution that looks similar to that of the Parisi solution,
our arguments show that it should be interpreted within the context of
possibility (d) above, i.e., dictatorial multiplicity,
rather than as confirmation of the SK picture.

To summarize our results, we have ruled out non-self-averaging in an
extremely large class of disordered systems, which include
short-ranged and probably all non-infinite-ranged spin glasses.
Non-self-averaging, and the other main consequences of the Parisi
solution, including
ultrametricity of pure state overlaps,
appear to be confined to mean field models.


\end{document}